\documentclass[letterpaper, paper,11pt]{AAS}		

\usepackage{bm}
\usepackage{float}
\usepackage{amsmath}
\usepackage{graphicx} 
\usepackage{subfigure}
\usepackage[colorlinks=true, pdfstartview=FitV, linkcolor=black, citecolor= black, urlcolor= black]{hyperref}
\usepackage{overcite}
\usepackage{footnpag}			      	
\newcommand{\bs}[1]{\boldsymbol{#1}}
\usepackage{xfrac}
\usepackage[outdir=./]{epstopdf}
\usepackage{bm}
\usepackage{soul}

\PaperNumber{23-205}

\title{Covariance Analysis of Attitude and Angular Rate Estimation using Accelerometers}

\author{Koya Yamamoto\thanks{Graduate Research Assistant, Aerospace Engineering Department, Texas A\&M University},  
Patrick Kelly\thanks{Graduate Research Assistant, Aerospace Engineering Department, Texas A\&M University},
Manoranjan Majji\thanks{Associate Professor, Director, LASR Laboratory,Aerospace Engineering Department, Texas A\&M University},\ 
and Felipe Guzm\'an\thanks{Associate Professor, Aerospace Engineering Department, Texas A\&M University}
}

\begin{document}

\maketitle{}

\begin{abstract}
    In this work a method for using accelerometers for the determination of angular velocity and acceleration is presented. Minimum sensor requirements and insights into how an array of accelerometers can be configured to maximize estimator performance are considered. The framework presented utilizes linear least squares to estimate functions that are quadratic in angular velocity. Simple methods for determining the sign of the spin axis and the linearized covariance approximation are presented and found to perform quite effectively when compared to results obtained by Monte Carlo. 
\end{abstract}

Accelerometers are a fundamental component in inertial measurement units (IMUs) and are offered in varying levels of cost and accuracy. In spaceborne applications they are used as both navigation equipment and scientific instruments. MEMS sensors are standard hardware on many low cost satellites \cite{cass2001mems}. State of the art, high precision acceleration measuring sensors, such as those used on the GRACE-FO mission are integrated into the data processing pipeline for gravity recovery \cite{kang2020grace}. The relative size, weight, and power requirements as well as overall cost of highly precise accelerometers continues to decrease with advances in optmechanical devices \cite{hines2020optomechanical,wisniewski2020optomechanical}. As access to highly accurate accelerometers expands it is natural that they may play a greater role in various aspects of spacecraft navigation and control.

Spacecraft attitude is determined through unit vector measurements, typically taken from star tracker observations. In between these star tracker observations (or in the circumstance that they are unavailable) the spacecraft attitude is propagated using angular rate measurements. These can be combined in a Kalman filter framework where the star tracker observations are used at the update step to correct the accumulated error over the propagation \cite{lefferts1982,zanetti2009norm,crassidis2007survey}. This combination provides flexibility and redundancy in the attitude estimation system and improves the overall performance. Accelerometers measure the contribution of non-inertial forces on the spacecraft motion. Because accelerometers are generally fixed in the body reference frame (BRF) these include the kinematic accelerations resulting from the rotation of the spacecraft. As a result accelerometer measurements can be combined with or used as a replacement for gyro measurements, permitting improved filter performance \cite{chheda06}, gyro/accelerometer calibration \cite{yu16}, or for gyro-free attitude estimation \cite{hamley22}.

In this work a method for the estimation of angular rates using measurements from an array of triaxial accelerometers and linear least squares is presented. Starting with the kinematic relations a simplified measurement model is developed. Using this measurement model a batch estimator employing $N$ sensors is given, along with the associated estimate error. Insights into the performance of this method as a function of the accelerometer configuration are discussed. The validity of the linear covariance analysis used to obtain the results for angular rate errors is assessed by way of Monte Carlo analysis.


\section{Kinematics and Measurement Model}
To estimate the angular rates of the body frame, an array of triaxial accelerometers are distributed at multiple locations on the spacecraft. These accelerometers are fixed in the body frame and are located with respect to the center of mass by the position vector $\bs{a}_i$, with components:
\begin{align}
    \bs{a}_i &= x_i \hat{b}_1 + y_i \hat{b}_2 + z_i \hat{b}_3  
\end{align}
The rotational motion of the body frame with respect to the inertial frame is then described using the angular velocity vector ($\bs{\omega} = \omega_1\hat{b}_1 + \omega_2\hat{b}_2 + \omega_3\hat{b}_3$). Using this, the kinematics can be developed to give the following inertial velocity and acceleration of the point $\bs{a}_i$:
\begin{gather}
    \dot{\bs{a}}_i = \dot{\bs{r}}_c +  \bs{\omega} \times \bs{a}_i = [\bs{\omega} \times]\bs{a}_i \\  \ddot{\bs{a}}_i = \ddot{\bs{r}}_c + \dot{\bs{\omega}} \times \bs{a}_i + \bs{\omega} \times(\bs{\omega} \times \bs{a}_i) = \ddot{\bs{r}}_c + \big([\dot{\bs{\omega}} \times] + [\bs{\omega}\times]^2\big)\bs{a}_i
\end{gather}
where $[\bs{\omega} \times]$ is the cross product matrix for the vector $\bs{\omega}$. Because the transnational acceleration of the center of mass is independent of the sensor configuration, the contribution of the spacecraft's transnational motion can be removed by differencing two acceleration terms. For this purpose a basepoint sensor located at $\bs{a}_0$ is designated, giving the relative accelerations: $\bs{g}_i = \ddot{\bs{a}}_i - \ddot{\bs{a}}_0$. Using this convention the acceleration can be expressed as a linear function of $\bs{a}_{i0} = \bs{a}_i-\bs{a}_0$:
\begin{figure}
    \centering
    \includegraphics[width=0.7\textwidth]{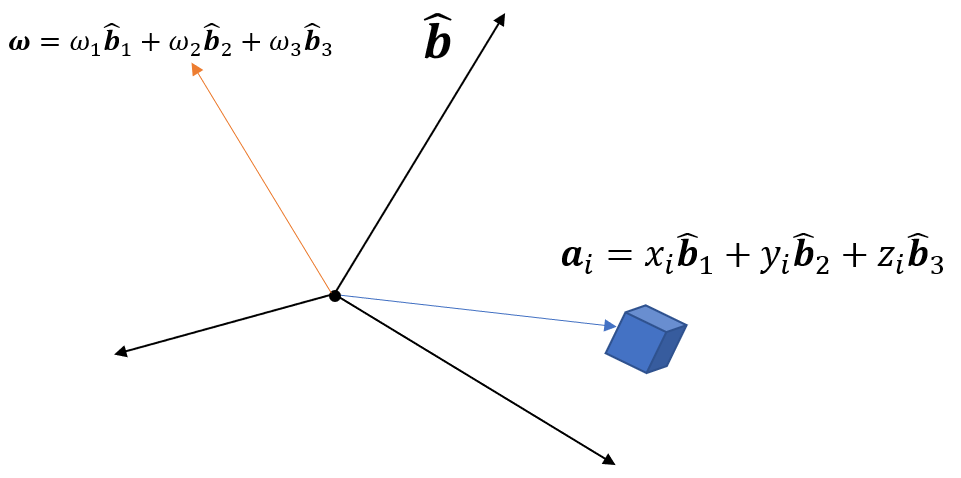}
    \caption{Accelerometer $i$ is fixed in the body frame and positioned by $\bs{a}_i$. The acceleration measured by this sensor is the sum of the translational acceleration and the gyroscopic acceleration induced by the angular velocity vector $\bs{\omega}$.}
    \label{fig:AccelArray}
\end{figure}

\begin{gather}
    \bs{g}_i = \mathbf{A}(\bs{\omega},\dot{\bs{\omega}})(\bs{a}_i - \bs{a}_0) =
    \begin{bmatrix} -(\omega_2^2+\omega_3^2) & \omega_1\omega_2-\dot{\omega}_3 & \omega_1\omega_3+\dot{\omega}_2 \\
    \omega_1\omega_2+\dot{\omega}_3 & -(\omega_1^2+\omega_3^2) & \omega_2\omega_3-\dot{\omega}_1 \\ 
    \omega_1\omega_3-\dot{\omega}_2 & \omega_2\omega_3+\dot{\omega}_1 & -(\omega_1^2+\omega_2^2)
    \end{bmatrix}
    \begin{bmatrix} x_{i0} \\ y_{i0} \\ z_{i0} \end{bmatrix}
\end{gather}
This provides a simple expression for the idealized measurement model of the rotating accelerometer. It is clear from the above expression that the components of $\bs{\omega}$ do not appear linearly in the measurement model. For this reason, linear least squares (LLS) is applied to estimate the components of $\mathbf{A}$ rather than the angular velocity vector itself.

\section{Linear Least Squares Solution}
Using the idealized measurement model developed in the previous section a procedure for the estimation of the components of $\mathbf{A}$ can be developed. The accelerometer measurements are given by:
\begin{gather}
    \tilde{\bs{g}}_i = \bs{g}_i + \bs{\nu}_i
\end{gather}
where $\bs{\nu}_i$ is zero-mean Gaussian noise with covariance: $\mathbf{R}_i = E[\bs{\nu}_i\bs{\nu}_i^T] = \sigma^2_{g,i}\mathbf{I}_{3\times 3}$. By this convention all accelerometer channels are taken to be independent. For an array made up of $N+1$ accelerometers the observation equation can be expressed as a linear system of the vectorized components of $\bs{\alpha} = \texttt{vec}(\mathbf{A})$:
\begin{gather}
    \begin{bmatrix} x_{10} & y_{10} & z_{10} & 0 & 0 & 0 & 0 & 0 & 0 \\ 
    0 & 0 & 0 & x_{10} & y_{10} & z_{10} & 0 & 0 & 0 \\ 0 & 0 & 0 & 0 & 0 & 0 & x_{10} & y_{10} & z_{10} \\
    x_{20} & y_{20} & z_{20} & 0 & 0 & 0 & 0 & 0 & 0 \\ 0 & 0 & 0 & x_{20} & y_{20} & z_{20} & 0 & 0 & 0 \\ 0 & 0 & 0 & 0 & 0 & 0 & x_{20} & y_{20} & z_{20} \\ \vdots & \vdots & \vdots & \vdots & \vdots & \vdots & \vdots & \vdots & \vdots \\ x_{N0} & y_{N0} & z_{N0} & 0 & 0 & 0 & 0 & 0 & 0 \\ 0 & 0 & 0 & x_{N0} & y_{N0} & z_{N0} & 0 & 0 & 0 \\ 0 & 0 & 0 & 0 & 0 & 0 & x_{N0} & y_{N0} & z_{N0} \end{bmatrix}
    \begin{bmatrix} A_{11} \\ A_{12} \\ A_{13} \\ A_{21} \\ A_{22} \\ A_{23} \\ A_{31} \\ A_{32} \\ A_{33} \end{bmatrix} = \begin{bmatrix} \tilde{\bs{g}}_1 \\ \tilde{\bs{g}}_2 \\ \vdots \\ \tilde{\bs{g}}_N
    \end{bmatrix}_{3N\times 1} \label{eq:NDim_LLS}
\end{gather}
If this system is expressed in matrix-vector form as: $\mathbf{H}_N\bs{\alpha} = \tilde{\bs{g}}$, then the LLS result can be applied directly to obtain the minimum variance estimate\cite{crassidis04} for $\mathbf{A}$:
\begin{gather}
    \hat{\bs{\alpha}} = \big[ \mathbf{H}_N^T\mathbf{R}^{-1}\mathbf{H_N} \big]^{-1} \mathbf{H}_N^T \mathbf{R}^{-1}\; \Tilde{\bs{g}}
\end{gather}
where $\mathbf{R} = \texttt{diag}(\mathbf{R}_1, \mathbf{R}_2, ... \mathbf{R}_N)$. Upon investigation it is clear that the system given in Equation \ref{eq:NDim_LLS} requires a minimum of 9 measurements to be uniquely defined. It follows that $N\geq 3$ measurements are required to provide a valid solution to the problem, meaning a total of 4 triaxial accelerometers are required including the basepoint sensor. Moreover, it is required that no combination of any 3 accelerometers are colinear in order to obtain a system with full rank. With this in mind, consider the $N=3$ case:
\begin{gather}
    \begin{bmatrix} x_{10} & y_{10} & z_{10} & 0 & 0 & 0 & 0 & 0 & 0 \\ 
    0 & 0 & 0 & x_{10} & y_{10} & z_{10} & 0 & 0 & 0 \\ 0 & 0 & 0 & 0 & 0 & 0 & x_{10} & y_{10} & z_{10} \\
    x_{20} & y_{20} & z_{20} & 0 & 0 & 0 & 0 & 0 & 0 \\ 0 & 0 & 0 & x_{20} & y_{20} & z_{20} & 0 & 0 & 0 \\ 0 & 0 & 0 & 0 & 0 & 0 & x_{20} & y_{20} & z_{20} \\ x_{30} & y_{30} & z_{30} & 0 & 0 & 0 & 0 & 0 & 0 \\ 0 & 0 & 0 & x_{30} & y_{30} & z_{30} & 0 & 0 & 0 \\ 0 & 0 & 0 & 0 & 0 & 0 & x_{30} & y_{30} & z_{30} \end{bmatrix}
    \begin{bmatrix} A_{11} \\ A_{12} \\ A_{13} \\ A_{21} \\ A_{22} \\ A_{23} \\ A_{31} \\ A_{32} \\ A_{33} \end{bmatrix} = \begin{bmatrix} \tilde{\bs{g}}_1 \\ \tilde{\bs{g}}_2 \\ \tilde{\bs{g}}_3
    \end{bmatrix}_{9\times 1}
\end{gather}
For accelerometers with equivalent measurement noise ($\sigma_{g}\equiv \sigma_{g,1} = \sigma_{g,2} = \sigma_{g,3}$), the LLS result is expressed simply as a matrix inversion problem:
\begin{gather}
    \hat{\bs{\alpha}} = \mathbf{H}_3^{-1}\tilde{\bs{g}}
\end{gather}
where the constant coefficient matrix inverse can be expressed using the Kronecker product ($\otimes$):
\begin{gather}
    \mathbf{H}_3^{-1} = \frac{1}{(\bs{a}_{10}\times\bs{a}_{20})\cdot\bs{a}_{30}}\begin{bmatrix} \mathbf{I}_3 \otimes (\bs{a}_{20}\times\bs{a}_{30}) & & \mathbf{I}_{30} \otimes (\bs{a}_{30}\times\bs{a}_{10}) & & \mathbf{I}_3 \otimes (\bs{a}_{10}\times\bs{a}_{20}) \end{bmatrix}
\end{gather}
Thus, the solution for the estimate of $\mathbf{A}$ is given simply as a matrix product. It is inferred that this compact form of the least-squares operator can be extended for $N>3$ with the use of the $n$-dimensional cross product; however this is left unresolved for the time being. The covariance of the estimate $\hat{\bs{\alpha}}$ for this simplified case can then be expressed as:
\begin{gather}
    \mathbf{P}_{\bs{\alpha}} = E[(\bs{\alpha}-\hat{\bs{\alpha}})(\bs{\alpha}-\hat{\bs{\alpha}})^T] = \begin{bmatrix} \mathbf{M}_{3\times 3} & \mathbf{0}_{3\times 3} & \mathbf{0}_{3\times 3} \\ \mathbf{0}_{3\times 3} & \mathbf{M}_{3\times 3} & \mathbf{0}_{3\times 3} \\ \mathbf{0}_{3\times 3} & \mathbf{0}_{3\times 3} & \mathbf{M}_{3\times 3}
    \end{bmatrix} \\
    \mathbf{M} = \frac{\sigma_g^2}{\big((\bs{y}\times\bs{z})\cdot\bs{x}\big)^2}\begin{bmatrix} 
    (\bs{y}\times\bs{z})^T(\bs{y}\times\bs{z}) & (\bs{y}\times\bs{z})^T(\bs{z}\times\bs{x}) & (\bs{y}\times\bs{z})^T(\bs{x}\times\bs{y}) \\
    (\bs{z}\times\bs{x})^T(\bs{y}\times\bs{z}) & (\bs{z}\times\bs{x})^T(\bs{z}\times\bs{x}) & (\bs{z}\times\bs{x})^T(\bs{x}\times\bs{y}) \\
    (\bs{x}\times\bs{y})^T(\bs{y}\times\bs{z}) & (\bs{x}\times\bs{y})^T(\bs{z}\times\bs{x}) & (\bs{x}\times\bs{y})^T(\bs{x}\times\bs{y}) 
    \end{bmatrix}
\end{gather}
\begin{align*}
    &\bs{x} = \begin{bmatrix} x_{10} & x_{20} & x_{30} \end{bmatrix}^T
    &\bs{y} = \begin{bmatrix} y_{10} & y_{20} & y_{30} \end{bmatrix}^T
    & &\bs{z} = \begin{bmatrix} z_{10} & z_{20} & z_{30} \end{bmatrix}^T
\end{align*}
Noting that the triple product can be expressed as $(\bs{y}\times\bs{z})\cdot\bs{x} = (\bs{z}\times\bs{x})\cdot\bs{y} = (\bs{x}\times\bs{y})\cdot\bs{z}$. Alternatively, the matrix $\mathbf{M}$ can be expressed as:
\begin{gather}
    \mathbf{M} = \frac{\sigma_g^2}{|\mathbf{K}|} \mathbf{K}\mathbf{K}^T \\
    \mathbf{K} = \begin{bmatrix}(\bs{a}_{20}\times\bs{a}_{30}) & (\bs{a}_{30}\times\bs{a}_{10}) & (\bs{a}_{10}\times\bs{a}_{20})\end{bmatrix} = \begin{bmatrix}(\bs{y}\times\bs{z})^T \\ (\bs{z}\times\bs{x})^T \\ (\bs{x}\times\bs{y})^T\end{bmatrix}
\end{gather}
This compact form for the covariance provides a direct relation between the configuration of the accelerometer array and the estimators performance. The covariance is scaled based on the determinant of $\mathbf{K}$, which is the volume of the parallelepiped formed by the column vectors. There are two direct consequences of this. First is that larger $\|\bs{a}_{i0}\|$ leads to smaller covariance. Second is that the for a fixed distance from the basepoint sensor, the smallest covariance is achieved when all $\bs{a}_{i0}$ are orthogonal. These covariance expressions all assume that $\sigma_g$ is equivalent for all accelerometers. If this is not the case the expressions become a bit more complicated, requiring the definition of $\mathbf{M}_i$ which is a function of \textit{each} of the terms $\sigma_{g,i}$. Leaving the specific form of each $\mathbf{M}_i$ asside, the block structure of the covariance is retained and these two properties hold. This fact will be useful for generalizing results moving forward.
\section{Angular Velocity Estimate and Linear Covariance Analysis}
Using the estimated matrix $\hat{\bs{\alpha}} \longrightarrow \hat{\mathbf{A}}$, the components of $\bs{\omega}$ and $\dot{\bs{\omega}}$ can be obtained. To do this, a transformation between $\mathbf{A}$ and $\{\bs{\omega},\dot{\bs{\omega}}\}$ must be defined. It is useful to note that the matrix is symmetric in $\bs{\omega}$ and skew-symmetric in $\dot{\bs{\omega}}$. For this reason, if we take the difference between the corresponding off-diagonal elements, the angular acceleration can be obtained:
\begin{subequations}
\begin{align}
    \dot{\omega}_1 &= \sfrac{1}{2}(A_{32}-A_{23})\\
    \dot{\omega}_2 &= \sfrac{1}{2}(A_{13}-A_{31}) \\
    \dot{\omega}_3 &= \sfrac{1}{2}(A_{21}-A_{12})
\end{align}
\end{subequations}
An additional consequence of $\dot{\bs{\omega}}$ appearing skew-symmetric is that the main diagonal is composed only of angular velocity terms. Using this the components of $\bs{\omega}$ can be expressed as a function of the main-diagonal entries:
\begin{subequations}
\begin{align}
    \omega_1^2 &= \sfrac{1}{2}(A_{11}-A_{22}-A_{33}) \\
    \omega_2^2 &= \sfrac{1}{2}(A_{22}-A_{11}-A_{33}) \\
    \omega_3^2 &= \sfrac{1}{2}(A_{33}-A_{11}-A_{22})
\end{align}
\end{subequations}
Another way to compute this is to define the symmetric matrix: $\mathbf{S} = \sfrac{1}{2}(\mathbf{A}^T+\mathbf{A})$. The eigenvalues of $\mathbf{S}$ are a single $\lambda = 0$  and a repeated $\lambda = -\omega^2$. The eigenvector associated with $\lambda = 0$ ($\bs{\zeta}^{(0)}$) is the axis of rotation.\par
Note that in both of these forms, the sign of $\bs{\omega}$ is not strictly defined due to the sign ambiguity resulting from the square root or the eigenvector. This is a result of the fact that the matrix $\mathbf{A}$ is identical for both $\bs{\omega}$ and $-\bs{\omega}$, as the angular velocity is a quadratic term in the kinematic acceleration expression. In short: the spin \textit{axis} can be defined, but the direction of rotation about this axis cannot be. To deal with this ambiguity some a-priori knowledge of $\bs{\omega}$ must be provided. This combined with the conservative assumption that the angular velocity vector is not rotating $>90^\circ$ in one sampling interval allows for a simple procedure to resolve the sign ambiguity:
\begin{align*}
    &\bs{\zeta}^{(0)} \cdot \hat{\bs{\omega}}(t_{k-1}) > 0    &\longrightarrow  & &\hat{\bs{\omega}}(t_k) = \bs{\zeta}^{(0)} \\
    &\bs{\zeta}^{(0)} \cdot \hat{\bs{\omega}}(t_{k-1}) < 0    &\longrightarrow  & &\hat{\bs{\omega}}(t_k) = -\bs{\zeta}^{(0)}
\end{align*}

The angular accelerations are expressed as a linear combination of the components of $\bs{\alpha}$, which allows for the direct transformation of the covariance obtained from the LLS solution. This gives the following:
\begin{subequations}
\begin{align}
    \sigma_{\dot{\omega}_1}^2 &= \frac{1}{4}(P_{88} + P_{66} - 2P_{68}) \\
    \sigma_{\dot{\omega}_2}^2 &= \frac{1}{4}(P_{33} + P_{77} - 2P_{37}) \\
    \sigma_{\dot{\omega}_3}^2 &= \frac{1}{4}(P_{22} + P_{44} - 2P_{24})
\end{align}
\end{subequations}
where $P_{ij}$ refer to the entries of $\mathbf{P}_{\bs{\alpha}}$. If the measurements are independent then $P_{68}=P_{37}=P_{24} = 0$, which reduces the expressions further. This form can be used if the measurement noise parameters are different for each accelerometer, but if they are equivalent the computed expression for $\mathbf{P}_{\bs{\alpha}}$ can be substituted to provide:
\begin{subequations}
\begin{align}
    \sigma_{\dot{\omega}_1}^2 &= \frac{\sigma_g^2}{4|\mathbf{K}|^2}\big[(\bs{z}\times\bs{x})^T(\bs{z}\times\bs{x}) + (\bs{x}\times\bs{y})^T(\bs{x}\times\bs{y})\big] \\
    \sigma_{\dot{\omega}_2}^2 &= \frac{\sigma_g^2}{4|\mathbf{K}|^2}\big[(\bs{x}\times\bs{y})^T(\bs{x}\times\bs{y}) + (\bs{y}\times\bs{z})^T(\bs{y}\times\bs{z})\big] \\
    \sigma_{\dot{\omega}_3}^2 &= \frac{\sigma_g^2}{4|\mathbf{K}|^2}\big[(\bs{z}\times\bs{x})^T(\bs{z}\times\bs{x}) + (\bs{y}\times\bs{z})^T(\bs{y}\times\bs{z})\big]
\end{align}
\end{subequations}
Unlike the previous case, the components of angular velocity are not expressed linearly as a function of $\bs{\alpha}$. As a result a linearized transformation is conducted in its place:
\begin{gather}
    \frac{\partial \bs{\omega}}{\partial \bs{\alpha}} = \frac{1}{4}\begin{bmatrix}
    \sfrac{1}{\omega_1} & 0 & 0   & 0 & -\sfrac{1}{\omega_1} & 0    & 0 & 0 & -\sfrac{1}{\omega_1} \\
    -\sfrac{1}{\omega_2} & 0 & 0   & 0 & \sfrac{1}{\omega_2} & 0    & 0 & 0 & -\sfrac{1}{\omega_2} \\
    -\sfrac{1}{\omega_3} & 0 & 0   & 0 & -\sfrac{1}{\omega_3} & 0    & 0 & 0 & \sfrac{1}{\omega_3}
    \end{bmatrix}
\end{gather}
\begin{gather}
    \mathbf{P}_{\bs{\omega}} = \frac{\partial \bs{\omega}}{\partial \bs{\alpha}} \mathbf{P}_{\bs{\alpha}} \frac{\partial \bs{\omega}}{\partial \bs{\alpha}}^T = \frac{1}{16} \begin{bmatrix}
    \frac{P_{11}+P_{55}+P_{99}}{\omega_1^2} & \frac{P_{99}-P_{11}-P_{55}}{\omega_1\omega_2} & \frac{P_{55}-P_{11}-P_{99}}{\omega_1\omega_3} \\ 
    \frac{P_{99}-P_{11}-P_{55}}{\omega_1\omega_2} & \frac{P_{11}+P_{55}+P_{99}}{\omega_1\omega_2} & \frac{P_{11}-P_{55}-P_{99}}{\omega_2\omega_3} \\ 
    \frac{P_{55}-P_{11}-P_{99}}{\omega_1\omega_3} & \frac{P_{11}-P_{55}-P_{99}}{\omega_2\omega_3} & \frac{P_{11}+P_{55}+P_{99}}{\omega_3^2}
    \end{bmatrix} \label{eq:LinOmegaCov}
\end{gather}
As in the previous case, this expression will hold for differing noise parameters across the different accelerometers. If the values of $\sigma_{g,i}$ are all taken to be equal then the diagonal elements are given as:
\begin{align}
    &\sigma_{\omega_1}^2 = \frac{s}{\omega_1^2}
    &\sigma_{\omega_2}^2 = \frac{s}{\omega_2^2} 
    & &\sigma_{\omega_3}^2 = \frac{s}{\omega_3^2} 
\end{align}
\begin{gather}
    s = \frac{\sigma_g^2}{16|\mathbf{K}|^2} \Big[(\bs{y}\times\bs{z})^T(\bs{y}\times\bs{z}) + (\bs{z}\times\bs{x})^T(\bs{z}\times\bs{x}) + (\bs{x}\times\bs{y})^T(\bs{x}\times\bs{y}) \Big]
\end{gather}

The same observations can be made with regard to the covariance of $\bs{\omega}$ and $\dot{\bs{\omega}}$ as with $\bs{\alpha}$. It is observed that the distance of each sensor from $\bs{a}_0$ is inversely proportional to the elements of the covariance matrix, and the covariance is minimized when the sensors are orthogonal.  To assess the validity of this linearized approximation of the angular velocity covariance, a numerical demonstration will now be provided.

\section{Monte Carlo Simulation}
\begin{figure}[H]
    \centering
    \includegraphics[width=\textwidth]{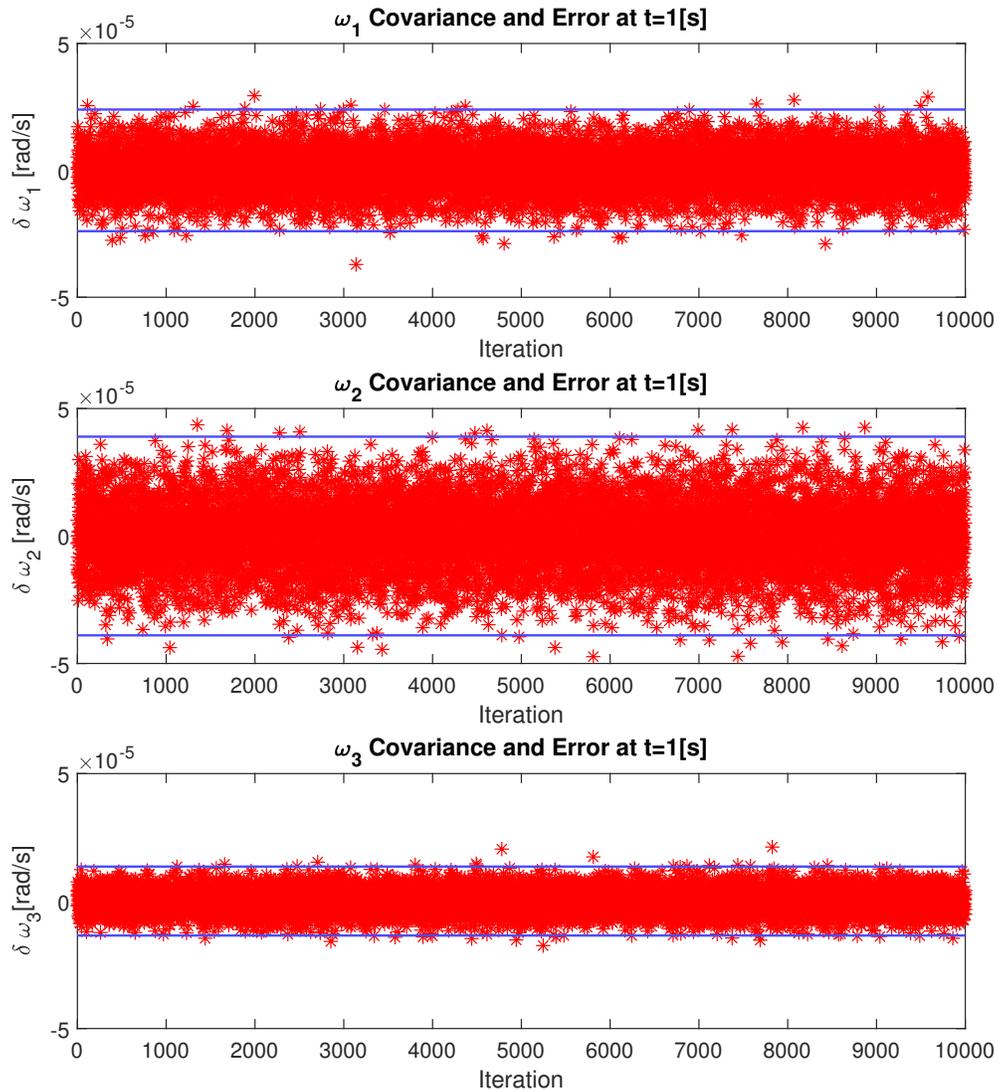}
    \caption{\(\omega\) covariance and error}
    \label{fig:o1}
\end{figure}

\begin{figure}[H]
    \centering
    \includegraphics[width=\textwidth]{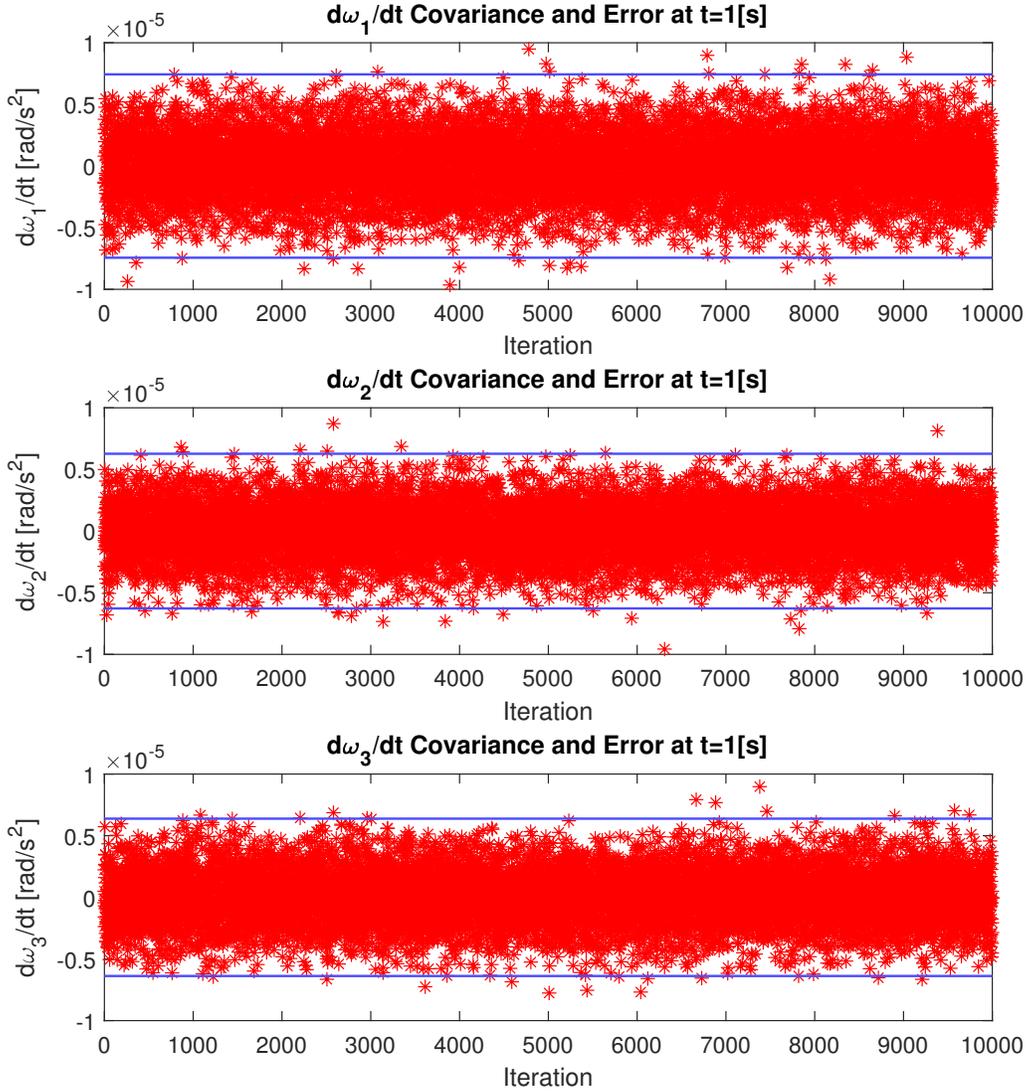}
    \caption{\(d\omega\) covariance and error}
    \label{fig:do1}
\end{figure}

In this section, the result of monte-carlo analysis is showed with 10,000 iteration, varying the measurement noise in each iteration. The iteration results are taken when the value of angular velocity and the angular acceleration is as follows
\begin{align}
    \omega = [-0.1716\;\;-0.1056\;\;-0.3043]^T \\ 
    \dot{\omega} = [0.0901 \;\; 0.0225 \;\; -0.0266]^T \\
    I = diag([14.3850,95.7408,55.5078]).
\end{align}
The configuration of the accelerometer has been determined randomly such that they each lie on the sphere, but are not orthogonal. \\
The results are showed in the below figures, and the blue line is the 3-\(\sigma\) bound, red dots are the errors. Figure \ref{fig:o1} is the result for the angular velocity and figure \ref{fig:do1} is the result for the angular acceleration.

Looking at the results overall, mostly the error are within 3-\(\sigma\) bounds and it seems the angular velocity, angular acceleration estimation is working.
Each angular velocity, acceleration covariance and error has slightly different values. This is because the configuration of the accelerometers as well as the initial angular velocities are different.

\begin{figure}[h]
    \centering
    \includegraphics[width=0.8\textwidth]{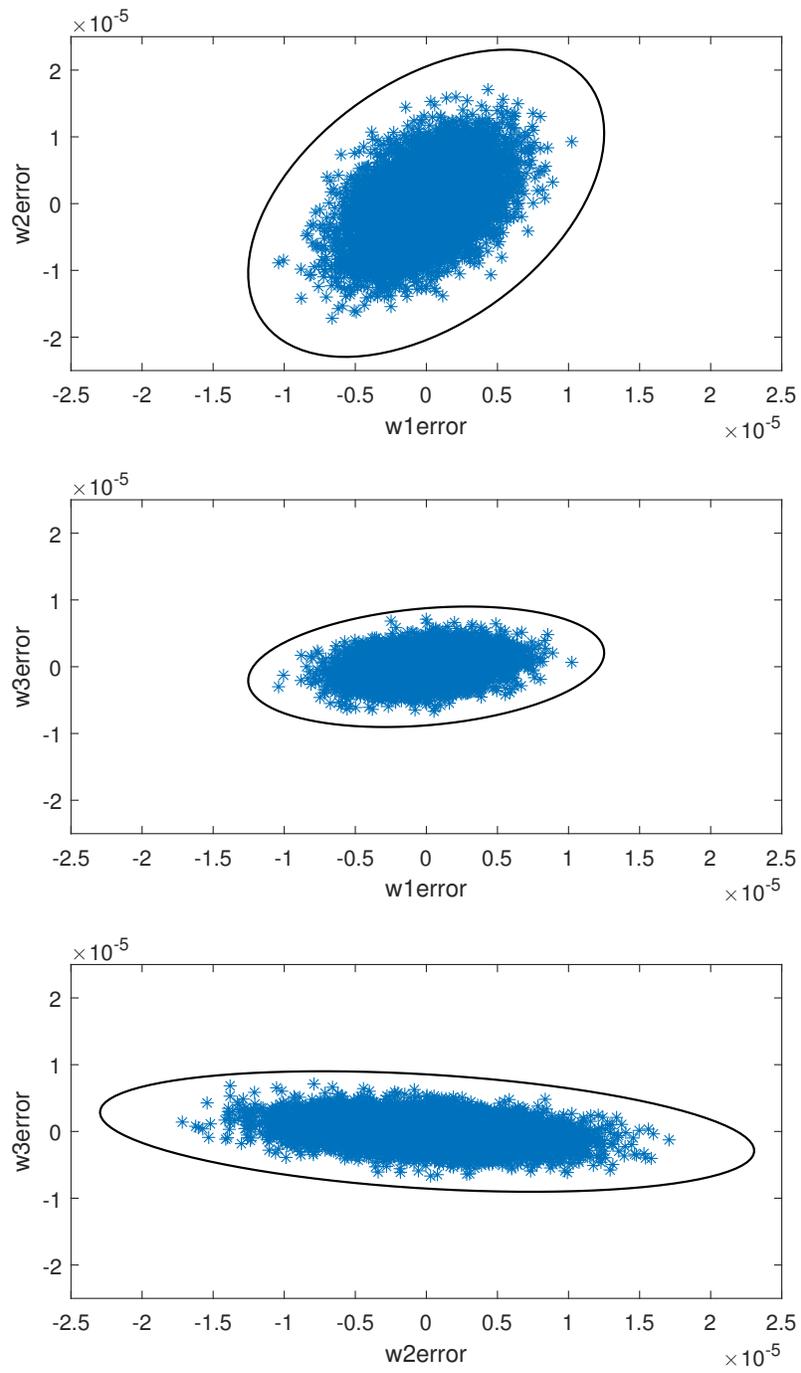}
    \caption{Covariance Ellipsoid and Angular Velocity Error}
    \label{fig:do3}
\end{figure}

Next each error of angular velocity and the covariance ellipsoid is showed in figure \ref{fig:do3}. All the error plots are within the covariance ellipsoid and it seems the result is decent. We can also observe from this figures that error of the \(\omega_2\) is larger than error of \(\omega_1\) or \(\omega_3\). The linearized approximation of the angular velocity covariance seems to be working here, as well.   

\section{Conclusion}
The estimation of angular velocity, acceleration using the accelerometers measurement were conducted using the linear least square method. The covariance analysis and the configuration of the accelerometers were also considered. 

The estimation of $\mathbf{A}$ matrix was first conducted, and then each angular velocity, acceleration was computed from the measurement model relation. Angular acceleration was expressed linearly by the components of estimated $\mathbf{A}$ matrix and the covariance of angular acceleration was determined by the direct transformation of the covariance obtained from the linear least square. For the angular velocity, linearized transformation was necessary, since the angular velocity was not expressed linearly as a function of components of $\mathbf{A}$ matrix. The validity of this linearized approximation and the estimation method was showed by the numerical demonstration, Monte Carlo simulation. The measurement error was varied in the Monte Carlo simulation with 10,000 iterations, and the error of the estimation was mostly within the 3-\(\sigma\) boundary. From here it is concluded that the angular velocity, acceleration estimation is working. Also the covariance ellipsoid for the angular velocity showed the validity of the linearized approximation for the angular velocity covariance. 

The configuration of the accelerometers were also considered from the linear least square covariance equation. The further accelerometers are located from the base accelerometer, the covariance is smaller. Moreover, smallest covariance is achieved when all the accelerometers are orthogonal. 


\bibliographystyle{AAS_publication}   
\bibliography{References}   

\end{document}